\documentclass[twocolumn]{aastex631}

\newcommand{\mbh}{${\cal M}_{\rm BH}$}

\newcommand{\kps}{${\rm km\,s}^{-1}$}
\newcommand{\ppxf}{{\sc ppxf}}
\newcommand{\vlos}{$v_{\rm los}$}
\newcommand{\slos}{$\sigma_{\rm los}$}

\usepackage{xcolor}

\shorttitle{A Supermassive Black Hole in a Diminutive Ultra-compact Dwarf Galaxy}
\shortauthors{Taylor et al.}

\begin{document}

\title{A Supermassive Black Hole in a Diminutive Ultra-compact Dwarf Galaxy Discovered with JWST/NIRSpec+IFU}

\author[0000-0003-3009-4928]{Matthew A.\ Taylor}
\affiliation{University of Calgary,
2500 University Drive NW,
Calgary Alberta T2N 1N4, CANADA}

\author[0000-0002-1584-2281]{Behzad Tahmasebzadeh}
\affiliation{Department of Astronomy, 
University of Michigan, 
1085 S. University Ave., 
Ann Arbor, MI 48109, USA}

\author[0009-0006-7485-7463]{Solveig Thompson}
\affiliation{University of Calgary,
2500 University Drive NW,
Calgary Alberta T2N 1N4, CANADA}

\author[0000-0002-5038-9267]{Eugene Vasiliev}
\affiliation{Institute of Astronomy, 
Madingley Road, 
Cambridge, CB3 0HA, UK}
\affiliation{University of Surrey, Guildford, GU2 7XH, UK}

\author[0000-0002-6257-2341]{Monica Valluri}
\affiliation{Department of Astronomy, 
University of Michigan, 1085 S. University Ave., 
Ann Arbor, MI 48109, USA}

\author[0000-0003-4867-0022]{Michael J.\ Drinkwater}
\affiliation{School of Mathematics and Physics, 
The University of Queensland, 
Brisbane, QLD 4072, Australia}

\author[0000-0003-1184-8114]{Patrick C\^ot\'e}
\affiliation{National Research Council of Canada, 
Herzberg Astronomy and Astrophysics Program, 
5071 West Saanich Road, 
Victoria, BC, V9E 2E7, Canada}

\author[0000-0002-8224-1128]{Laura Ferrarese}
\affiliation{National Research Council of Canada, 
Herzberg Astronomy and Astrophysics Program, 
5071 West Saanich Road, 
Victoria, BC, V9E 2E7, Canada}

\author[0000-0002-0363-4266]{Joel Roediger}
\affiliation{National Research Council of Canada, 
Herzberg Astronomy and Astrophysics Program, 
5071 West Saanich Road, 
Victoria, BC, V9E 2E7, Canada}

\author[0000-0002-1959-6946]{Holger Baumgardt}
\affiliation{School of Mathematics and Physics, 
The University of Queensland, 
Brisbane, QLD 4072, Australia}

\author[0000-0002-2816-5398]{Misty C.\ Bentz}
\affiliation{Department of Physics and Astronomy, Georgia State University, Atlanta, GA 30303, USA}

\author[0000-0002-8532-4025]{Kristen Dage}
\affiliation{International Centre for Radio Astronomy Research $-$ Curtin University, GPO Box U1987, Perth, WA 6845, Australia}

\author[0000-0002-2073-2781]{Eric W. Peng}
\affiliation{NSF NOIRLab, 950 N.\ Cherry Ave., Tucson, AZ 85719, USA}

\author[0009-0009-5509-4706]{Drew Lapeer}
\affiliation{Department of Astronomy, University of Michigan, 1085 S. University Ave., Ann Arbor, MI 48109, USA}
\affiliation{Department of Astronomy, University of Massachusetts, 710 North Pleasant Street, Amherst, MA 01003, USA}

\author[0000-0002-4718-3428]{Chengze Liu}
\affiliation{Department of Astronomy, School of Physics and Astronomy, and Shanghai Key Laboratory for Particle Physics and Cosmology, Shanghai Jiao Tong University, Shanghai 200240, People's Republic of China}

\author[0009-0004-8539-3516]{Zach Sumners}
\affiliation{University of Calgary, 2500 University Drive NW, Calgary Alberta T2N 1N4, CANADA}
\affiliation{Trottier Space Institute, McGill University, 3550 Rue University, Montreal QC H3A 0G4, Canada}

\author[0000-0002-3382-9021]{Kaixiang Wang}
\affiliation{Department of Astronomy, 
Peking University, 
Beijing 100871, People's Republic of China}
\affiliation{Kavli Institute for Astronomy and Astrophysics, 
Peking University, 
Beijing 100871, People's Republic of China}

\author[0000-0003-4703-7276]{Vivienne Baldassare}
\affiliation{Department of Physics and Astronomy, Washington State University, Pullman, WA 99163, USA}

\author[0000-0002-5213-3548]{John P.\ Blakeslee}
\affiliation{NSF NOIRLab, 950 N.\ Cherry Ave., Tucson, AZ 85719, USA}

\author[0000-0001-6333-599X]{Youkyung Ko}
\affiliation{Korea Astronomy and Space Science Institute, 776 Daedeok-daero, Yuseong-Gu, Daejeon 34055, Republic of Korea}

\author[0000-0003-1428-5775]{Tyrone E.\ Woods}
\affiliation{Department of Physics and Astronomy,
University of Manitoba,
30A Sifton Road,
Winnipeg, Manitoba, R3T 2N2 Canada}



\begin{abstract}

The integral-field unit mode of the Near-Infrared Spectrograph (NIRSpec+IFU) mounted on the James Webb Space Telescope has now enabled kinematic studies of smaller and less massive compact stellar systems in which to search for central massive black holes (BHs) than ever before. We present here the first such detection using NIRSpec+IFU in its highest resolution ($R\sim2700$) mode. We report the detection of a central black hole with mass {\bf \mbh$=2.1\pm1.1\times10^6\,M_\odot$} ($1\sigma$ uncertainties) in UCD736 orbiting within the Virgo galaxy cluster. Schwarzschild modeling of the 1D kinematic profile rules out a zero-mass central black hole at the $3\sigma$ level; however, two other independent modeling approaches fail to rule out a zero-mass black hole at $>1\sigma$ significance. The presence of such a massive BH strongly argues against a globular cluster origin of this UCD, and rather suggests a tidally stripped formation route from a former $\gtrsim10^9\,M_\odot$ dwarf galaxy host. This represents the detection of a BH in the most compact ($r_h\approx15\,{\rm pc}$) stellar system to date, with a \mbh\ corresponding to {\bf $\sim 8$ percent} of the system's stellar mass, roughly in line with previously reported UCD BH detections and comparable to the BH detected in the compact elliptical galaxy NGC4486B.

\end{abstract}



\section{Introduction} \label{sec:intro}
Around twenty years ago, a purported new type of stellar system was discovered orbiting within the dense environments of the Virgo and Fornax galaxy clusters \citep{hil99, dri00, has05}. These objects, quickly termed ``ultra-compact dwarf galaxies'' (UCDs), exhibit luminosities and half-light radii ranging from $-14\,{\rm mag} \lesssim M_V \lesssim -11\,{\rm mag}$ and $10-50$\,pc. At the time, it was unclear whether this new class of objects was galactic in origin (e.g., the final remains of nucleated dwarf galaxies stripped of their diffuse retinue of stars) or merely represented the high-luminosity tail of the globular cluster luminosity function.

Massive black holes (BHs) have been known to exist in high-mass UCD analogues, such as compact elliptical galaxies (cEs). In particular, the Virgo cluster cE VCC1297 (NGC\,4486B) was reported to harbor a black hole with a mass of \mbh$\approx 6\times10^8\,M_\odot$ \citep{kor97}. Using JWST NIRSpec/IFU data from the same program used to obtain data used in the current paper we have estimated the BH in this galaxy to be $3.6\pm0.6\times10^6\,M_\odot$ at $3\sigma$ confidence \citep{tah25}. In the local volume, searches for black holes in cE have yielded high-likelihood detections in NGC\,221 (M\,32) and NGC\,404, which contain black holes with masses of approximately \mbh$\approx 3\times10^6\,M_\odot$ and $\sim5\times10^5\,M_\odot$, respectively \citep{ben96, ver02, set10, dav20}.

The first black hole definitively discovered in a UCD was reported by \cite{set14} in M60-UCD1, located near the giant elliptical galaxy M60 within the Virgo cluster. This UCD was found to host a supermassive black hole (SMBH) with \mbh$=2.1^{+1.4}_{-0.7}\times10^7\,M_\odot$, corresponding to about 15 percent of M60-UCD1's stellar mass. Since this groundbreaking discovery, several other massive black holes have been detected in other UCDs, including three in the Virgo galaxy cluster. Two of these—M59-UCD3 and M59cO—are located within the M59/M60 Virgo C subcluster complex, with black hole masses of \mbh$=4.2\times10^6\,M_\odot$ and $5.8\times10^6\,M_\odot$, respectively \citep{ahn17, ahn18}. A fourth Virgo UCD confirmed to host a SMBH of \mbh$=4.4\times10^6\,M_\odot$ lies near the giant cD galaxy M87 \citep[VUCD3;][]{ahn17}. Beyond these, only a single other UCD in the Fornax galaxy cluster has been shown to harbor a massive BH, with \mbh$=3.3\times10^6\,M_\odot$ reported in the center of Fornax UCD3 \citep{afa18}, while similar searches in two of the most luminous compact stellar systems in Centaurus A have only yielded upper limits of $10^{5-6}\,M_\odot$ \citep{vog18}.

The subject of this work is UCD736, which is a small \citep[$r_h=14.96\pm0.47\,{\rm pc}$;][]{liu20} UCD located in the Virgo cluster, with a total luminosity of $M_V=-12.51$\,mag, placing it at the lower end of the UCD luminosity range. Based on multi-band ground-based photometry, it has a total stellar mass ${\cal M}_\star=10^{7.4}\,M_\odot$, making it among the densest such systems known \citep{liu15}. UCDs that are previously known to harbor black holes are listed in Table\,\ref{tbl:css_bhs} where we note that UCD736 stands out as both the smallest and least luminous specimen. Previous black hole detections have generally been restricted to UCDs lying at the higher range of sizes and/or luminosities, or in the much more luminous cE systems. Understanding the demographics of massive black holes residing in fainter, smaller, and lower mass UCDs is thus of significant interest to determine the fraction of UCDs that are of galactic or star cluster origins with implications on massive BH seeding mechanisms in the early universe \citep[e.g.,][]{gre20}.

In what follows, \S\ref{sec:obs} describes the two space-based datasets that this study analyzes, followed by a description of the derived 1D kinematics in \S\ref{sec:kin}. In \S\ref{sec:bh_model} we discuss three different approaches to modeling the derived kinematic profiles and summarize our findings in \S\ref{sec:disc}. Throughout this work, we assume a distance to UCD736 equal to the mean of the Virgo cluster \citep[$16.5\pm1.1$\,Mpc;][]{mei07}.


\begin{figure*}
\plotone{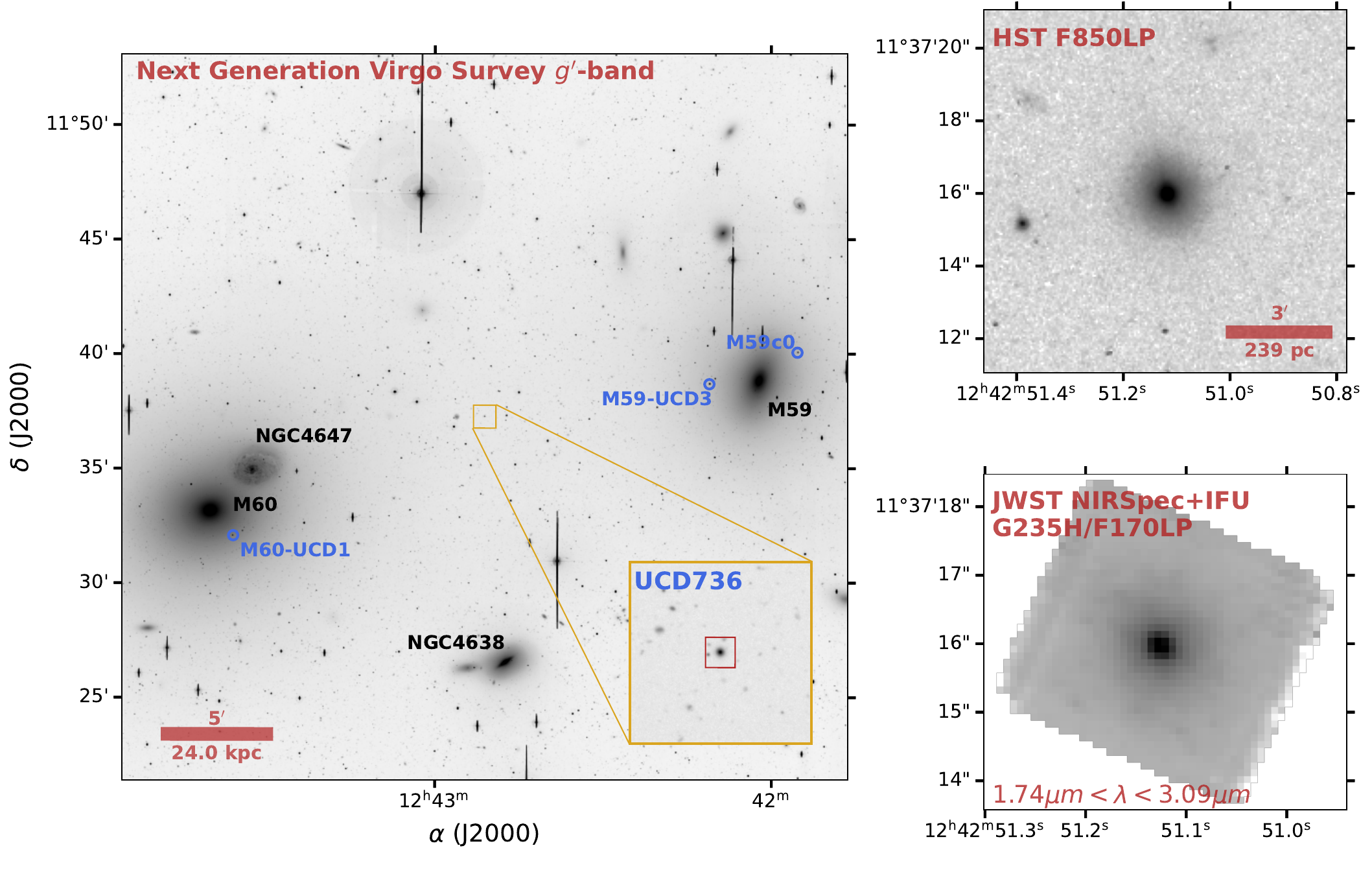}
\caption{The M59/M60 complex with UCD736 and SMBH-hosting UCDs of previous studies. The main panel shows an NGVS $g'$-band image with the location of UCD736 between M59 and M60 indicated by the yellow zoom-in inset. The top right-hand panel shows UCD736 imaged with HST's ACS/WFC in the F850LP filter used for our light-profile modeling, while the bottom panel shows a JWST/NIRSpec+IFU white-light image reconstructed from the $1.74\,\mu m<\lambda<3.03\,\mu m$ wavelength slices. Physical scale bars are presented in the main and HST panels for context.}
\label{fig:ucd736_im}
\end{figure*}

\begin{deluxetable*}{lrrclcrc}
\tablecaption{Summary of black hole detections in compact stellar systems\label{tbl:css_bhs}}
\tabletypesize{\scriptsize}
\tablehead{
\colhead{ID} & \colhead{$\alpha$} & \colhead{$\delta$} & \colhead{$M_V$} & \colhead{$r_h$} & \colhead{$\log{\cal M}_\star$} & \colhead{\mbh} &   \colhead{Reference} \\
\colhead{} & \colhead{(deg)} & \colhead{(deg)} & \colhead{(mag)} &
\colhead{(pc)} & \colhead{($\times10^6\,M_\odot$)} & \colhead{($\times10^6\,M_\odot$)}   & \colhead{}
}
\startdata
M60-UCD1    & 190.8999 &   11.5346 & $-14.14$ & $19.58\pm0.37$ & $8.1$  & $21^{+14}_{-7}$     &   \citep{liu20,set14} \\
M59-UCD3    & 190.5460 &   11.6448 & $-14.74$ & $25.01\pm0.27$ & $8.4$  & $4.2^{+2.1}_{-1.7}$ &   \citep{liu20,ahn18} \\
M59c0       & 190.4806 &   11.6677 & $-13.56$ & $36.40\pm0.21$ & $7.9$  & $5.8^{+2.5}_{-2.8}$ &   \citep{liu20,ahn17} \\  
VUCD3       & 187.7391 &   12.4290 & $-12.69$ & $20.08\pm0.46$ & $7.5$  & $4.4^{+2.5}_{-3.0}$ &   \citep{liu20,ahn17} \\
Fornax UCD3 &  54.7254 & $-35.5602$& $-13.33$ & $86.50\pm6.2$  & $7.9$   & $3.3^{+1.4}_{-1.2}$ &   \citep{evs08,afa18} \\
\hline
UCD736      & 190.7130 &   11.6211 & $-12.51$ & $14.96\pm0.47$ & $7.4$  & $2.1\pm1.1$         & (\citealt{liu20}, this work) 
\enddata
\tablecomments{General data for compact stellar systems with high likelihood massive black hole detections in their cores. Column (1) shows system IDs used in this work, columns (2-3) show on-sky coordinates of the systems, while the next four columns indicate absolute $V$-band magnitudes, projected half-light radii, stellar masses, and reported black hole masses in that order. The final column indicates data references where the first listing refers to columns (4), (5), and (6), and the second refers to column (7).}
\end{deluxetable*}

\section{Observations and Data Processing} \label{sec:obs}
Figure\,\ref{fig:ucd736_im} shows the location of UCD736 in the Virgo cluster, where the main panel shows its projected location nearly equidistant between the giant galaxies M60 and M59 on a $g'$ band image from the {\it Next Generation Virgo Survey} \citep{fer12}. Its projected separation from either of the giant galaxies is $\sim1500''$, suggesting a physical separation of $\sim160\,{\rm kpc}$, unlike the three other UCDs that have black hole detections indicated by the blue circles that are projected much closer to their host galaxies. The upper and lower right-hand panels show UCD736 in a {\it Hubble Space Telescope} (HST) image, and a reconstructed white-light image from the {\it James Webb Space Telescope}'s (JWST) Near-infrared Spectrograph integral-field unit \citep[NIRSpec+IFU;][]{bok22,jak22}, respectively.

Spectroscopic observations were made using the JWST on 18 January 2023 as part of Cycle 1 operations (PropID: 2576; PI: M.\ Taylor)\footnote{All data used in this paper can be found in MAST: \dataset[https://doi.org/10.17909/hg5f-q394]{https://doi.org/10.17909/hg5f-q394}} using JWST/NIRSpec+IFU. UCD736 was observed in the highest resolution setup using the G235H/F170LP grating/filter combination, resulting in datacubes measuring $\sim3\arcsec$ on a side, with $\sim0.1\arcsec$ spaxels corresponding to $\sim8$\,pc at our assumed distance. In IFU mode, each spaxel of the image contains spectra at resolution $R\sim2700$, corresponding to velocity FWHM $\Delta v_{r}\approx47\,$\kps\ in the $1.7\la\lambda/\mu m\la3.2$ wavelength range. The datacube was constructed from four individual exposures with a total integration time of $\sim3000$\,s, using a four-point dither pattern with sub-slice offsets $0.025''$ to improve spatial PSF sampling and used the NRSIRS2RAPID readout mode.

Datacubes were constructed using the STScI JWST/NIRSpec+IFU pipeline v1.14.1 using CRDS context 1227. During the Stage 1 initial reduction routines, ``snowball'' flagging was enabled to detect significant cosmic ray events with a rejection threshold of $4\sigma$ requiring at least 10 contiguous pixels to register as a cosmic ray event. For any positive detections, extra pixels surrounding the affected area were flagged as unusable. After detector-level data preparation, individually dithered data cubes underwent Stage 2 processing, including WCS information assignment, flat field correction, and wavelength calibration. Upon completion of this stage, individually resampled 3D data cubes were combined in Stage 3 processing into the final data cubes. Prior to Stage 3 processing, we included additional cleaning routines to increase the overall quality of the data cubes and reduce the number of spectral spikes and artifacts that we found to be imprinted on the data if the default routines were applied. These extra steps included flagging of pixels affected by unreliable regions of flat exposures, checking for saturated pixels in individual frames, or other calibration effects that result in individual bad pixels. Additionally, a check for pixels affected by failed open shutters on the NIRSpec microshutter assembly was performed with affected pixels flagged.

To maximize the accuracy of our light-profile deprojection used for the determination of potential black holes (see \S\ref{sec:bh_model}), we obtained supplementary imaging data with the HST. Images were obtained on 03 February 2022 with the Advanced Camera for Surveys Wide Field Channel (ACS/WFC) that has a pixel scale of $\sim0.05\arcsec\,{\rm pix}^{-1}$, or $\lesssim4$\,pc at UCD736's assumed distance; a factor $\gtrsim2$ improvement in the modeling of light profiles over the white image of NIRSpec + IFU alone. Specifically, we use a 1220 second integration taken in F850LP filter (PropID: 16882; PI: M.\ Taylor), that is sensitive to the age and metallicity properties expected of compact stellar systems such as UCDs. We used science-ready data products produced by the default MAST pipeline.


\section{Kinematic Measurements\label{sec:kin}}
The left panel of Figure\,\ref{fig:ppxf} shows the $S/N$ per spectral resolution element of each individual IFU spaxel based on the error spectrum provided by the STScI pipeline. The central brightest pixel shows $S/N\sim80$ with $S/N$ falling outside the core pixel to $\sim20$ at $\sim3r_h$. To increase spectral $S/N$ and minimize kinematic uncertainties, we radially bin the datacube into six summed annuli with widths of a single pixel, centered on the pixel with the highest $S/N$. In each azimuthally averaged bin we sum the spectra and add the corresponding uncertainties reported in the error spectrum in quadrature. The right-hand panels of Figure\,\ref{fig:ppxf} show example 1D spectra (black lines) corresponding to the single central pixel (top; $S/N=77$) and the outermost bin of 32 summed pixels (bottom; $S/N\approx55$) that were used in the BH modeling (see \S\,\ref{sec:bh_model}).

\subsection{Penalized Pixel Fitting \label{sec:ppxf}}
\begin{figure*}
\plotone{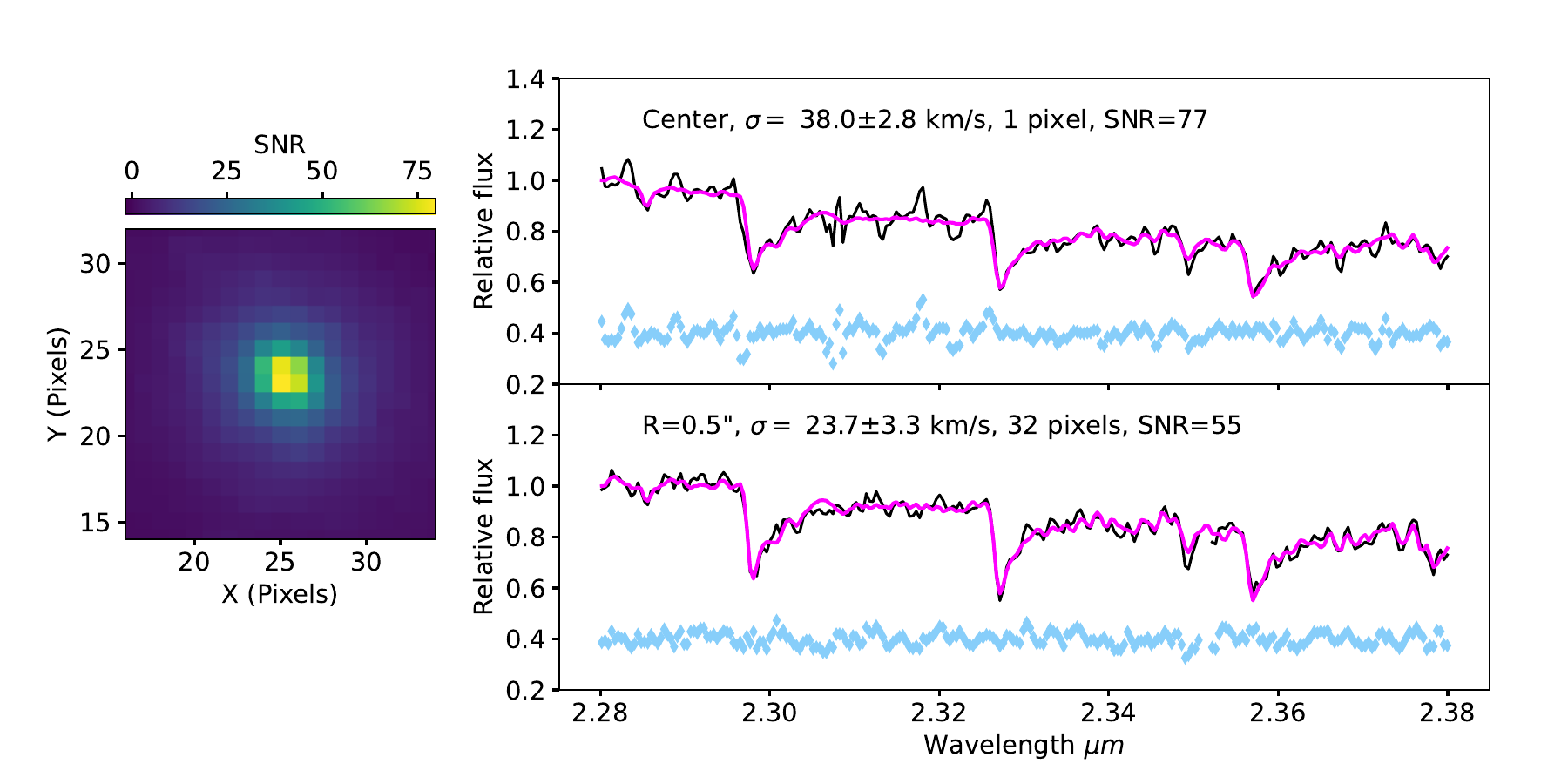} 
\caption{Signal to noise ratio map and \ppxf\ fits to central pixel and radius $0.5\arcsec$ of UCD736. The left-panel shows the signal-to-noise ratio per resolution element for each pixel considered in the NIRSpec/IFU field of view where we find a central maximum of $\gtrsim75$ with a rapid decline to zero beyond the extent of UCD736. Pixels are azimuthally binned to increase signal-to-noise prior to determining the kinematic profile. The right panels show {\sc ppxf} fits to the center-most pixel (top), with data shown in black, kinematic fit in purple, and residuals by blue points. Below is a similar fit to the azimuthally averaged pixels centred at a projected distance of $0.5\arcsec$ from the center. In both panels we indicate the number of IFU pixels contributing to the spectrum, and measured signal-to-noise ratio.}
\label{fig:ppxf}
\end{figure*}

The individual azimuthally summed 1D spectra are used to derive the stellar kinematics of the integrated light, namely the line-of-sight velocity distribution (LOSVD), using penalized pixel fitting software \citep[\ppxf;][]{cap03}. \ppxf\ parametrically recovers the internal stellar kinematics, particularly the line-of-sight radial velocity (\vlos) and velocity dispersion (\slos), by representing the LOSVD as a Gauss-Hermite series.
The code convolves a library of stellar spectral templates with a parametrized LOSVD to create a model spectrum with the resolution of the JWST spectra.
We used a library of $\sim600$ high-resolution synthetic spectra from the PHOENIX library \citep{hus13} covering a range of stellar parameters: 
$3800 \leq T_{\rm eff} \leq 6000$ K, 
$-1.5 \leq {\rm [Fe/H]} \leq 1.0 $ dex, 
$-0.2 \leq [\alpha/{\rm Fe}] \leq 2.0 $ dex, 
$0.5 \leq \log g \leq 2.5$.
Before modeling the kinematics, we truncate both the observed and stellar template spectra to the wavelength range $2.20-2.43 \,\mu{\rm m}$ to derive the stellar kinematics from the strong CO-bandheads near $2.29-2.38 \, \mu{\rm m}$ (see right-hand panel of Figure\,\ref{fig:ppxf}).

The initial LOSVD parameters input into \ppxf\ are the literature $537\,{\rm km\,s}^{-1}$ value for \vlos\ \citep{liu20}, and $40\,{\rm km\,s}^{-1}$ for \slos. The optimal $v_0$, $\sigma_0$, and optional Hermite moments $h_3$ and $h_4$ are then determined by minimizing $\chi^2$ between the model and the observed galaxy spectrum. Using an iterative process \ppxf\ derives a penalty function from the integrated square deviation of the line profile from the best fitting Gaussian, which allows for higher-order details to be recovered when $S/N$ is high, but biases the fit towards a Gaussian when $S/N$ is low. We find that a 4th degree additive polynomial is sufficient to recover the stellar kinematics from the spectra.

Measurement of \vlos\ from azimuthally averaged 1D bins smooths over any rotational signature, but tests on the 2D datacube indicate that this rotation has a maximum peak-to-valley amplitude of $\lesssim10$\,\kps, with negligible rotation in the core region where the dynamical influence of any BH present would be the strongest. As a result, we do not expect this averaging to have any influence on our main results. From the integrated spectrum of all spaxels summed within $3r_h$ we find that \vlos$=519.8\pm0.7$\kps, falling within the uncertainties of the single \vlos$=537\pm30$\,\kps\ reported in the literature \citep{zha15}, noting that this value was based on a lower $R\sim1280$ spectrum.

Uncertainties for the kinematics are estimated via bootstrapping. Random Gaussian noise is added to each spectral pixel centred on the value of the flux at each wavelength, with a width corresponding to the uncertainty report in the binned error spectrum before being re-fit with \ppxf. We repeat this process 1\,000 times and adopt the $1\sigma$ spread in each parameter's results as our measured uncertainties. As one would expect, the accuracy of the kinematic fits is best toward the highest $S/N$ central bins ranging from uncertainties on \vlos\ and \slos\ of $\sim1-4$\,\kps\ and $\sim2-8$\,\kps, respectively (see Figure\,\ref{fig:bh_results}, middle panel). We find a similar trend in the uncertainties reported for $h_4$, which are indicated in the bottom panel of Figure\,\ref{fig:bh_results}.

The right-hand panels of Figure\,\ref{fig:ppxf} show example spectral fits (magenta lines) to the CO-bandhead in the $2.28\lesssim\lambda/\mu m\lesssim2.38$ wavelength region corresponding to the central pixel (top) and outermost annulus used in our analysis (bottom) from which kinematics were extracted. The fits by \ppxf\ to the spectra are good, and show a decrease from a central value of \slos$=38.0\pm2.8$\,\kps\ to a minimum value of \slos$=23.7\pm3.3$\kps\ in the outer bin.


We note that while the NIRSpec/G235H line-spread function varies by a factor of two across the entire spectral range, in the wavelength range we observe near the CO bandheads, the variation is only 5\% ($R=2615$ to 2740) so that any discrepancy in matching to the line-spread function is $\lesssim2$\,\kps, well within our recovered uncertainties. Still, it should be noted that the nominal velocity FWHM at $R\sim2\,700$ corresponds to $v_{\rm FWHM}\sim47$\,\kps, such that our recovered kinematics are all below this limit. To test the reliability of our measured \slos, we perform simulations based on a K3III PHOENIX stellar spectrum. We first broaden the simulated spectrum to the recovered kinematics for each radial bin, smooth it to the resolution of the NIRSpec/IFU line-spread function, and add Gaussian noise to degrade the spectrum to $30\leq S/N\leq 100$ in steps of $\Delta S/N=10$. We then create ten realizations of the simulated spectrum at each $S/N$ and measure the recovered kinematics with {\sc ppxf}.

From these experiments, we can reliably recover \slos$\gtrsim25$\,\kps\ for $S/N\gtrsim50-60$. However, the spread in recovered \slos\ at the $S/N$ of our outermost two bins is $\sim10$\,\kps, which we incorporate by adding $5$\,\kps\ in quadrature to our Monte Carlo derived uncertainties. Additionally, we find an offset in the recovered kinematics in the outer two bins such that the recovered kinematics are systematically $\sim5$\,\kps\ higher than the input values at similar $S/N$ to our data. We report the results of testing this potential bias in the Appendix.


\section{Black Hole Mass Estimation\label{sec:bh_model}}
\subsection{Instrument Resolution and Light Profile Modeling \label{sec:lightprof}}
We use the instrument resolutions to compare our models to observations. We parameterize the NIRSpec+IFU PSF as two circular Gaussians derived from stars observed with the same setup as this program \citep{ben25}, with $\sigma=0.06\arcsec$ and $\sigma=0.14\arcsec$, each normalized to contribute $71\%$ and $29\%$ of the PSF, respectively. We parameterize the HST F850LP PSF as three circular Gaussians with $\sigma=0.02\arcsec, \, 0.07\arcsec,$ and $ 0.3\arcsec$, each normalized to contribute $78.2\%, \, 21.6\%,$ and $0.2\%$ of the PSF, respectively.

We measure the visible light profile of UCD736 from the HST F850LP image using multi-Gaussian expansion \citep[MGE;][]{cap02}. MGE models the intrinsic surface brightness as a series of super-imposed 2D elliptical Gaussians \citep{ems94} that are convolved with the instrumental HST PSF prior to being fitted to match the observed image. The unconstrained MGE fits were very close to circular, so we restricted the fits to circular Gaussians to model UCD736 as a spherical system. 

Upon construction of our HST-based light profile, we then use it in three independent modeling methods to determine the \mbh\ of any potential black hole in the system. Specifically we model the system using Schwarzschild's dynamical modeling, Jeans Anisotropic modeling, and an approach using distribution functions as described in the following. We do not include a dark matter component in any of these models, as we assume that such a component would be stripped away upon passage through the host Virgo cluster potential \citep[e.g.,][]{fra11, tol11, str13}. Under this approach, the following analysis leads to a conservative lower limit to any central SMBH, given that adding a dark matter halo would actually increase the inferred black hole mass under the commonly used assumption of a constant stellar M/L \citep{she10}.

\subsection{Schwarzschild Dynamical Modeling} \label{sec:schw}
We utilize the Schwarzschild orbit superposition modeling technique \citep{sch79} as implemented in the FORSTAND code \citep{vas20}, which is included in the AGAMA stellar-dynamics toolbox \citep{vas19}. This framework has been previously successful in constructing dynamical models and used to determine black hole masses in two galaxies \citep{rob21,mer23} and used in extensive ``mock'' testing of UCD-like systems to determine limits on black hole masses that could realistically be recovered in Virgo Cluster dwarf galaxies with our observational setup \citep{tah24}.

A total gravitational potential is constructed that includes the contributions from the stars (related to density by Poisson's equation), and a black hole modeled as a Plummer potential with a fixed-scale radius of $a=10^{-4}$ kpc.  Since we do not consider a dark matter component for UCD736, we are left with $M/L$ and \mbh\ as the two free parameters allowed to vary in these models.

For every set of model parameters, we build an orbit library by integrating $N_{\rm orb}=$ 20,000 orbits over 100 dynamical periods within the specified potential. The LOSVD of each orbit is convolved with the PSF and recorded in each radial bin. The initial conditions for these orbits are sampled randomly.

The different orbit libraries are constructed on a grid of models that span a range of $0<$\mbh/M$_\odot<10^{7.2}$. Orbit weights are determined by minimizing the deviation between model and observed kinematics parameterized by Gauss-Hermite moments ($h_2$ and $h_4$) with {\it fixed} \slos\ as discussed in section 2.6 of \cite{vas20}. We compute the $\chi^{2}$, which assesses the fit relative to the observational constraints (\slos, $h_4$) and their corresponding uncertainties. Consequently, each orbit library is utilized multiple times to explore a range of $M/L$ values by multiplying the velocities by $\sqrt{M/L}$ in multiplicative steps of 0.01 until the minimum of $\chi^{2}$ is found and bracketed from both ends.

\subsection{Jeans Anisotropic modeling}
We estimate the black hole mass from the observed velocity dispersion using Jeans anisotropic modeling, as implemented in the jampy Python package (JAM; \citealt{cap08}). JAM identifies solutions to the Jeans equations where the gravitational potential is constructed from the stellar mass distribution plus a central black hole modelled as a Gaussian of mass \mbh. The stellar mass distribution is found by deprojecting the aforementioned MGE light profile to three dimensions and assuming a constant mass-to-light ratio ($M/L$). As noted above, we do not include dark matter. The velocity anisotropy $\beta$ is assumed to be constant across the object. For the spherical model we adopt, the velocity ellipsoid is aligned with a spherical coordinate system \citep{cap20}. For a spherical system the models are independent of inclination, so there are three free parameters: \mbh, $M/L$, and $\beta$. The models are integrated along the line-of-sight before being convolved with the NIRSpec+IFU PSF to generate the modelled stellar velocity dispersion profile, $\sigma_{\star,m}$. We used Markov chain Monte Carlo sampling to fit all three free parameters, and then marginalized the distributions over velocity anisotropy ($\beta$) to compare models as a function of $M/L$ and \mbh.

\begin{deluxetable}{lccc}
\tablecaption{Black hole modeling results}\label{tbl:bh_mods}
\tablehead{
\colhead{Technique}  &   \colhead{${\cal M}_{BH}$}  &   \colhead{$(M/L)_{F850LP}$}  &   \colhead{$f_{BH}$}  \\
\colhead{}  &   \colhead{($\times10^6\,M_\odot$)} &   \colhead{($M_\odot/L_\odot$)} &   \colhead{(\%\ ${\cal M}_\star$)}
}
\startdata
Schwarszchild   &   {\bf $2.1\pm1.1$}   	           &  $1.45\pm 0.15$ &   {\bf $8.4\pm4.4$} \\
JAM            	 &   $1.8^{+1.4}_{-1.8}$       &  $1.60\pm 0.20$ &   $7.2^{+5.6}_{-7.2}$  \\
DF              	 &   $1.9^{+0.9}_{-1.9}$   	  &  $1.68\pm 0.20$ &   $7.6^{+3.6}_{-7.2}$ \\
\enddata
\tablecomments{The uncertainties are $\pm 1 \sigma$ statistical ranges; we do not include the small (7\%) systematic uncertainty in distance.}
\end{deluxetable}

\begin{figure}
\plotone{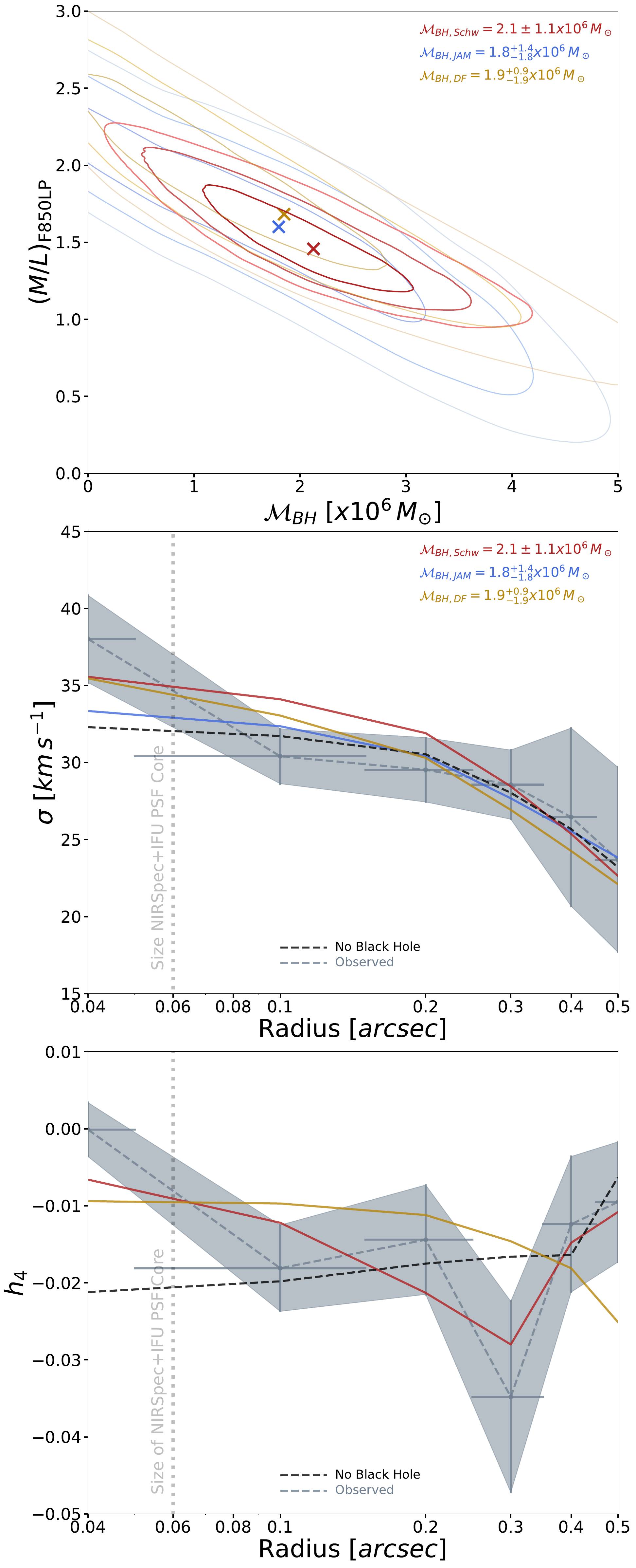} 
\caption{Summary of black hole modeling results. {\it Top panel}:\,$M/L$ vs.\ \mbh\ $\chi^2$ confidence contours of the JAM (blue), Schwarzschild (red), and DF (yellow) modeling results. Curves are shown at $1\sigma$, $2\sigma$, and $3\sigma$ confidence levels. {\it Middle panel}:\,\slos\ radial profile shown for the modeling results, with colors representing the same models as the top-panel. The observed profile is indicated by the grey dashed line, bounded by grey shading corresponding to the $1\sigma$ kinematic uncertainties, while the black dashed line shows the no black hole scenario in the Schwarszchild modeling. {\it Bottom panel}:\,$h_4$ radial profile shown for the DF (yellow) and Schwarzschild (red) modeling results, with the grey dashed line and shading representing the observed profile and $1\sigma$ uncertainties. Horizontal grey lines indicate the extent of the radial bins.}
\label{fig:bh_results}
\end{figure}

\subsection{Distribution function-based modeling\label{sec:df}}
As a third dynamical modeling approach, we use the method based on distribution functions (DFs) in the energy and angular momentum space ($E$--$L$). This approach has previously been used in applications to dwarf galaxies \citep[e.g.,][]{wil02, woj08}, and the present implementation has been extensively tested in \cite{rea21}. We use a spherical DF with a constant anisotropy $\beta$, which is determined using the \cite{cud91} generalization of the Eddington inversion method. In this approach, we have the same three free parameters as in the Jeans method ($\cal M_{\rm BH}$, $M/L$ and $\beta$), but unlike the latter, the DF can be guaranteed to be physically possible (nonnegative), and we can use higher-order Gauss--Hermite moments in comparing models to observations. For comparison with other modeling approaches in the $\cal M_{\rm BH}$--$M/L$ space, we picked the values of $\beta$ that gave the smallest $\chi^2$ for each combination of the other two parameters; these were usually in the range 0 (isotropic) to 0.2 (slight radial anisotropy).


\section{Discussion \& Summary} \label{sec:disc}
The upper panel of Figure\,\ref{fig:bh_results} shows the $\chi^2$ contours of $M/L$ and \mbh\ of the three techniques described above, corresponding to the confidence intervals $1\sigma$, $2\sigma$, and $3\sigma$. Red, blue, and yellow crosses indicate the models corresponding to the lowest $\chi^2$ for the Schwarzschild, JAM, and DF approaches, respectively, with \mbh\ summarized in Table\,\ref{tbl:bh_mods} alongside $M/L$ and their respective marginalized $1\sigma$ uncertainties. The best fit Schwarzchild model shows a stellar $M/L=1.45$, which is somewhat lower than what is predicted from fitting the optical photometry gathered by the NGVS \citep{fer12} to simple stellar population models \citep{bru03} that predict  $M/L\approx2.2$. If this slightly higher $M/L$ is adopted, then the Schwarzchild modeling would still predict a non-zero SMBH within the $3\sigma$ contours of Fig.\,\ref{fig:bh_results}. Moreover, if UCD\,736 has origins as a higher mass dwarf galaxy, then the assumption of an underlying simple stellar population may not be valid, so this comparison should be taken in that context.


We show in the middle panel of Figure\,\ref{fig:bh_results} the predicted \slos\ profiles by the colored lines overplotted on the determined kinematics of {\sc ppxf} shown by the gray dashed line bounded by uncertainty bars/shading. Horizontal grey bars indicate the spatial extent of each radial bin. The black dashed line corresponds to the no-black hole scenario predicted by the Schwarzschild modeling, which most closely follows the best-fit JAM model that shows the weakest constraints in the top panel.

We note here that the JAM and DF analyses can only formally provide $1\sigma$ upper limits on \mbh, which is likely due to the sensitivity of $h_4$ to the presence (or lack) of a central black hole. JAM does not include $h_4$, thus limiting its constraining power in marginal detections such as this, but the bottom panel of Figure\,\ref{fig:bh_results} shows the radial profile of $h_4$ measured by {\sc ppxf} compared with those predicted by the Schwarzschild and DF methods. While both of these methods can reproduce the $\sigma$-profile well, and each predict $\sigma_0$ within the observed uncertainties, it can be seen that the Schwarzschild model provides a much better fit to the $h_4$ profile, resulting in the tighter constraints seen in the upper panel. Moreover, the $h_4$ profile for the no black hole scenario predicted by Schwarzschild modeling can only reproduce the observed results in the outer bins, with significant discrepancy in the core.

Given the significantly smaller confidence intervals from Schwarzschild modeling, and that all three methods provide results within the associated $1\sigma$ confidence contours, we adopt the Schwarzschild model as a very likely positive detection of a {\bf \mbh$=2.1\pm1.1\times10^6\,M_\odot$} black hole residing in the core of UCD736. We further emphasize that this is the first attempt to measure a stellar dynamical BH mass applying three independent modeling methods to the same dataset. The good agreement between the $1\sigma$ confidence intervals of all three models bolsters our confidence in the estimated BH mass. The zero-mass black hole case is excluded at the $3\sigma$ level by the Schwarzschild contours in the top panel of Figure\,\ref{fig:bh_results}. This represents a black hole comprising {\bf $\sim4-13$ percent} of UCD736's ${\cal M}_\star=10^{7.4}\,M_\odot$ \citep{liu20}, roughly in line with the five previously detected UCD black holes \citep{set14, ahn17, ahn18,afa18}. Moreover, this detection is in line with theoretical predictions that black holes comprising $\gtrsim 1-10$ percent of a system's mass should be recoverable by JWST/NIRSpec+IFU using data of comparable quality to those analyzed here \citep{tah24}.

Taken at face value, UCD736 represents the fifth positive detection of a black hole in a UCD orbiting within the Virgo galaxy cluster \citep{set14,ahn17,ahn18}, joined by another black hole-dominated UCD in the Fornax galaxy cluster \citep{afa18}. However, UCD736 is the smallest, faintest, and least massive UCD in which such a massive black hole has been detected. Combined with previous results in more massive UCDs massive black holes are likely common across the UCD mass spectrum, and act as a strong discriminator between very massive/large globular clusters \citep[e.g,][]{fel02,fel05,kis06} or tidally stripped nucleated galaxy \citep[e.g.,][]{bek01,bek03,dri03} origins.

Comparing this result to the closest analogue in the Local Group --- namely $\omega$ Cen --- we note the claim of a lower-limit to a central massive black hole in that system of \mbh$\ga8200\,M_\odot$ (\citealt{hab24}, although see \citealt{ban25} for a counter-claim). Their lower-limit, even while taking our result as an upper-limit, implies a black hole in UCD\,736 up to $\sim100\times$ higher than $\omega$ Cen, despite a rather modest factor of $\sim7$ increase in $V$-band luminosity \citep{bau18}. While stronger constraints are needed for both systems, taken together these results may indicate that black hole formation becomes more efficient at galaxy masses $\ga10^9\,M_\odot$ or $\ga10^7\,M_\odot$ nucleus mass. Refining the current measurement further will be difficult, as improving either the spectral or spatial resolution using current ground-based instrumentation is either impossible or prohibitively expensive in terms of required observation time.

Like those UCDs reported before, we interpret the presence of a central supermassive black hole as evidence of a galactic origin for UCD736. Following the scaling relation of \cite{mcc13}, this black hole suggests a significantly more massive progenitor. Assuming mass is conserved during the stripping process that creates UCDs, a black hole mass of {\bf $2.1\times10^6\,M_\odot$} implies a progenitor with velocity dispersion $\sim90$\,\kps\ \citep[][their Figure\,1]{mcc13}. This corresponds to a present day Virgo member of $L_g\approx3\times10^9\,L_{\odot,g}$, or $M_\star\approx9\times10^9\,M_\odot$, comparable to or exceeding that of a typical dwarf elliptical galaxy or the bulge of a low-mass spiral galaxy. UCD736's relatively modest mass compared to other UCDs with positive black hole detections might be taken as significant, but recent work by \cite{wan23} shows that there is a full evolutionary sequence that efficiently transforms such dwarfs into UCD-like compact systems. In this light, UCD736 then represents the extension of this process to lower final mass and one may expect to continue detecting such black holes in the cores of UCDs with properties that encroach upon the globular cluster regime.

Given that the mean velocity dispersion of galaxies in the Virgo cluster of $\sim 650$km~s$^{-1}$ \citep{mclaugh99} is much larger than the internal velocity dispersion of UCD\,736, the galaxy has probably experienced much more significant stripping from the mean cluster tidal field \citep{Valluri1993} than from any individual galaxy. At its current distance  from the Virgo cluster center (875.5~kpc), assuming a NFW mass profile with parameters from \cite{mclaugh99}, the tidal radius of UCD\,736 is $\sim1.4$~kpc --- nearly 10 times larger than its current optical radius. This suggests that it is likely on an eccentric orbit that has previously taken it much closer to the cluster center, where it was heavily stripped. 

The location of UCD\,736 $\sim\!160$\,kpc from the nearest giant galaxy (M\,59) in Virgo makes it clear that UCDs with tidally stripped galactic origins need not be found exclusively in close proximity to giant galaxies as previous UCDs with BH detections have been\footnote{Although M60 is also at $\sim\!160$\,kpc from UCD\,736, the relative velocity between the two galaxies is over 500~km~s$^{-1}$, making tidal interaction unlikely.}. The relative line-of-sight velocity between  UCD\,736 and M\,59 is $\sim70$\,\kps\ implying that UCD~736 might be interacting tidally with M\,59. However, we find that the tidal radius of M\,59 (arising from the cluster tidal field) at its current distance from the cluster center is $\sim 110$~kpc implying that UCD\,736 is outside M\,59's tidal radius and is not currently bound to it. Still, the possibility of a previous interaction with M\,59 calls into question the validity of our assumed distance of $16.5$\,Mpc. M\,59 is $\sim10$ percent closer at $14.9$\,Mpc \citep{bla09}, which would decrease our assumed \mbh\ by a similar factor that remains within our uncertainties.

Regardless, this is the fourth such black hole detected in UCDs orbiting within the same Virgo sub-cluster, which all form an almost perfectly linear distribution spanning 100s of kpc in projection. This may be suggestive either that the M\,59/M\,60 complex is particularly effective at stripping (and not in fact due to the cluster tidal field) or that more examples might still be found around the M87 complex of the Virgo galaxy cluster.

\begin{acknowledgments}
We thank the anonymous referee for helpful and constructive comments that served to improve the original version of this manuscript. M.A.T.\ and S.T.\ acknowledge funds from the CSA (22JWGO1-07) and the University of Calgary graduate program. B.T.\ and M.V.\  acknowledge funding from Space Telescope Science Institute awards: JWST-GO-02567.002-A and HST-GO-16882.002-A. E.V.\ acknowledges support from an STFC Ernest Rutherford fellowship (ST/X004066/1). The work of E.W.P.\ and J.P.B.\ is supported by NOIRLab, which is managed by the Association of Universities for Research in Astronomy (AURA) under a cooperative agreement with the U.S. National Science Foundation. M.C.B. gratefully acknowledges support from the NSF through grant AST-2407802. Based on observations with the NASA/ESA Hubble Space Telescope obtained from the Barbara A.\ Mikulski Archive for Space Telescopes (MAST) at the Space Telescope Science Institute, which is operated by the Association of Universities for Research in Astronomy, Incorporated, under NASA contract NAS5-26555. Support for Program number GO-2567 was provided through a grant from the STScI under NASA contract NAS5-26555.

\end{acknowledgments}

%

\vspace{5mm}
\facilities{JWST(NIRSpec), HST(ACS/WFC)}


\software{astropy \citep{ast13},
          galfit \citep{pen10},
          jam \citep{cap20},
          matplotlib \citep{hun07},
          mge \citep{cap02},
          numpy \citep{van11},
          scipy \citep{vir20},
          ppxf \citep{cap03},
          agama \citep{vas19},
          forstand \citep{vas20}.
          }

\appendix
\section{Testing Potential Bias in Outermost Kinematic Bins}
Given the systematic offset by $\sim5$\,\kps\ of our recovered kinematics in the outermost two radial bins based on our mock spectrum (see \S\,\ref{sec:kin}), we perform a test to check on the potential effect on the recovered black hole mass. For this test, we decreased the velocity dispersion in the outer two bins by $5$\,\kps\ while also adding $5$\,\kps\ in quadrature to our Monte Carlo uncertainties. We find that a zero-mass black hole is still strongly constrained at the $3\sigma$ level, with an even more massive $3.37^{+1.30}_{-1.39}\times10^6\,M_\odot$ black hole recovered. This result is to be expected, as a lower dispersion requires a correspondingly lower stellar mass to explain it. However, given the assumption of a constant stellar mass-to-light ratio throughout the UCD, a higher mass black hole is then required to compensate for the lower stellar mass in the outer bins. With that said, the result of this experiment also predicts a much lower $M/L\approx0.9$ that is implausible for a UCD from a stellar population synthesis standpoint. For this reason, our reported $M_{BH}$ might be considered a conservative lower limit.

%



\end{document}